\documentclass[aps,pre,showpacs,amsmath,amssymb,amsfonts,superscriptaddress,reprint,10pt]{revtex4-1}

\usepackage{bm}
\usepackage{dsfont}
\usepackage{verbatim}
\usepackage{graphicx}

\usepackage[colorlinks=true,allcolors=blue,breaklinks=true]{hyperref}

\newcommand{\beq}{\begin{equation}}
\newcommand{\eeq}{\end{equation}}
\newcommand{\beqa}{\begin{eqnarray}}
\newcommand{\eeqa}{\end{eqnarray}}

\newcommand\rmd{{\mathrm{d}}}
\newcommand\rme{{\mathrm{e}}}
\newcommand\rmi{{\mathrm{i}}}

\newcommand{\RR}{{\mathds R}}
\newcommand{\CC}{{\mathds C}}

\newcommand{\one}{{\mathds 1}}

\DeclareMathOperator{\Tr}{Tr}

\usepackage{xcolor}

\begin{document}

\title{Relaxation timescales and prethermalisation in \texorpdfstring{$d$}{d}-dimensional long-range quantum spin models}

\author{Michael Kastner}
\email{kastner@sun.ac.za}
\affiliation{Institute of Theoretical Physics,  University of Stellenbosch, Stellenbosch 7600, South Africa}
\affiliation{National Institute for Theoretical Physics (NITheP), Stellenbosch 7600, South Africa}

\author{Mauritz van den Worm}
\affiliation{Institute of Theoretical Physics,  University of Stellenbosch, Stellenbosch 7600, South Africa}
\affiliation{National Institute for Theoretical Physics (NITheP), Stellenbosch 7600, South Africa}

\begin{abstract}
We report analytic results for the correlation functions of long-range quantum Ising models in arbitrary dimension. In particular, we focus on the long-time evolution and the relevant timescales on which correlations relax to their equilibrium values. By deriving upper bounds on the correlation functions in the large-system limit, we prove that a wide separation of timescales, accompanied by a pronounced prethermalisation plateau, occurs for sufficiently long-ranged interactions.
\end{abstract}

\maketitle

Empirically it is well established that, after a sufficiently long time, most physical many-body systems, whether isolated or coupled to an environment, will equilibrate. In many cases the equilibrium is well described by a Gibbs state, and this observation is at the basis of equilibrium statistical mechanics. An understanding of the microscopic process leading to thermalisation is, however, still incomplete. Recent experiments with cold atoms, ions, and molecules \cite{Kinoshita_etal06,Hofferberth_etal07,Struck_etal11,Islam_etal11,Trotzky_etal12,Britton_etal12,Islam_etal13,Yan_etal13} have sparked a revival of interest in questions related to the foundations of equilibrium statistical mechanics. Substantial progress, often based on typicality techniques, has been made on the theoretical side in the past few years. Results in various physical settings have been reported, proving that equilibration and/or thermalisation takes place for typical quantum systems of sufficient size \cite{Reimann08,Linden_etal09,
Goldstein_etal10,Reimann10,ReimannKastner12,UdudecWiebeEmerson13}.

While these results establish that equilibration and/or thermalisation will happen eventually, the time scale
of such a relaxation process remains unspecified. Only very recently has it become apparent that typicality
techniques can also be applied for analysing the time scales of equilibration. The results of these efforts,
while pioneering a promising approach, are not yet fully satisfactory, as they either over-
\cite{ShortFarrelly12} or---in a different setting---underestimate \cite{GoldsteinHaraTasaki} the timescales
by many orders of magnitude. These results indicate that physically realistic models or observables are not
``typical'' in the mathematical sense.

In this paper we approach the problem of equilibration time scales from a different angle, reporting the results of a model study. This work is a continuation and extension of a previous paper \cite{vdWorm_etal13} where exact, analytical expressions for equal-time two-point correlation functions have been computed for long-range interacting Ising models in a longitudinal magnetic field. While these results are very general and exact, the long-time asymptotics relevant for the relaxation to equilibrium is not at all obvious from the analytical expressions. Motivated by recent ion-trap experiments where the Ising model with long-range interactions can be realised, we have derived in \cite{vdWorm_etal13} upper bounds on the time-evolution of the various spin--spin correlation functions of the two-dimensional long-range Ising model on a triangular lattice. For the long-range Ising model on this specific lattice, the asymptotic long-time behaviour can be read off easily from the upper bounds. In the present paper 
we report generalisations of these upper bounds to arbitrary regular lattices and dimensionality. 

We consider long-ranged coupling constants $J_{i,j}\propto\left|i-j\right|^{-\alpha}$ decaying like a power law with the distance $\left|i-j\right|$ between lattice sites $i$ and $j$. The exponent $\alpha\geq0$ in some sense quantifies the range of the interactions, from infinite range in the case $\alpha=0$ to nearest-neighbour interactions in the limit $\alpha\to\infty$. The upper bounds derived in this paper are stretched or compressed exponentials in time for all values of $\alpha$. On a $d$-dimensional lattice and for $\alpha<d/2$, we find that some of the spin--spin correlation functions relax to their equilibrium values in a two-step process, governed by two widely separate time scales, while single-spin expectation values relax already on the faster of the two time scales. This kind of behaviour, characterised by a long-lived quasi-stationary state in which only some of the expectation values have already relaxed to their equilibrium value goes under the name of ``prethermalisation'' and has been 
discussed extensively in the past few years \cite{Berges_etal04,MoeckelKehrein08,EcksteinKollarWerner09,Gring_etal12,MarinoSilva12,AduSmith_etal13,GongDuan13,Marcuzzi_etal13}.

The stretched or compressed exponentials that upper-bound the spin--spin correlation functions do not only depend on the exponent $\alpha$, but they do so in a nonanalytic, transition-like manner: the long-time asymptotic behaviour of the spin--spin correlation functions switches from one kind of behaviour to a different functional form at the values $\alpha=d/2$ and $\alpha=d/2-1$. The first of these threshold values was already discussed in \cite{vdWorm_etal13} for the triangular lattice in two dimensions, and it is related to the occurrence of widely separated time scales and prethermalisation. The second threshold value, at $\alpha=d/2-1$, has not been described before, as it becomes relevant (i.e., occurs at a positive $\alpha$-value) only for lattice dimensions $d\geq3$.

\section{Long-range Ising model}
Consider a lattice $\Lambda$ consisting of $N$ sites, to each of which is assigned a spin-$1/2$ degree of freedom. Each spin is modelled by a Hilbert space $\mathcal{H}_i=\CC^2$, and the composite system is described by the Ising-type Hamiltonian 
\begin{equation}\label{e:H2}
H_\ell=-\frac{1}{2}\sum_{i\in\Lambda} \sum_{j\in\Lambda\backslash\{i\}}J_{i,j}\sigma_i^z\sigma_j^z-h\sum_{i\in\Lambda}\sigma_i^z,
\end{equation}
on the tensor product Hilbert space $\mathcal{H}=\bigotimes_{i\in\Lambda}\mathcal{H}_i$. The parameter
$h\in\RR$ denotes the magnitude of a homogeneous external magnetic field in the $z$ direction, and
$\sigma_i^z$ is the $z$-Pauli operator acting on the $i$th component of the tensor product space
$\mathcal{H}$. At this point the coupling strengths $J_{i,j}$ between lattice sites $i,j\in\Lambda$ are
arbitrary, but we will specialise to power law-decaying long-range interactions at a later stage.

As initial states we choose density operators $\rho_0$ that are diagonal matrices in the $\sigma^x$ tensor-product eigenbasis,
\begin{multline}\label{e:rho0}
\rho_0=\frac{1}{2^N}\biggl(\one+\sum_{i}\sigma_i^x\biggl(s_i^x + \sum_{j>i}\sigma_j^x\biggl(s_{ij}^{xx}\\
 + \sum_{k>j}\sigma_k^x\biggl(s_{ijk}^{xxx} + \sum_{l>k}\cdots\biggr)\biggr)\biggr)\biggr),
\end{multline}
where $\one$ denotes the identity operator on $\mathcal{H}$. The indices $i$, $j$, $k$, $l$ in \eqref{e:rho0} are summed over the lattice sites. This choice of $\rho_0$ has been exploited previously \cite{Emch66,Kastner11,Kastner12,vdWorm_etal13}, as it leads to particularly simple calculations and results, although generalisations are possible. Starting from $\rho_0$, exact analytic expressions have been reported in \cite{vdWorm_etal13} for the time-evolution of the various spin--spin correlation functions, e.g.
\begin{equation}\label{e:xx}
\begin{split}
 \langle \sigma^x_i \sigma^x_j \rangle (t)&:=\Tr \left(\rme^{\rmi H_\ell t} \sigma_i^x\sigma_j^x \rme^{-\rmi H_\ell t}\rho_0\right) \\
&= \langle \sigma^x_i \sigma^x_j \rangle(0)\left(P^{-}_{i,j}+P^{+}_{i,j}\right)
\end{split}\end{equation}
with
\begin{equation}\label{e:P}
 P^{\pm}_{i,j}=\frac{1}{2}\prod_{k\in\Lambda\setminus\{i,j\}} \cos\left[2\left(J_{k,i}\pm J_{k,j}\right)t\right],
\end{equation}
where we have set $h=0$ for simplicity. Other correlation functions either behave similarly (like $\langle \sigma^y_i \sigma^y_j \rangle$), or simpler (like $\langle \sigma^y_i \sigma^z_j \rangle$), or are constant in time (like $\langle \sigma^z_i \sigma^z_j \rangle$); see \cite{vdWorm_etal13} for details. In the following we restrict the presentation to the $xx$-correlation function given in \eqref{e:xx}, but similar techniques can be applied to other correlation functions. Upper bounds on one-point functions like $\langle \sigma^x_i \rangle$ have been reported in the Supplemental Material accompanying Ref.\ \cite{BachelardKastner13}.

\section {Upper bound on the correlations}

The expressions in \eqref{e:P} are quasi-periodic in $t$, and it is therefore precluded that $\langle
\sigma^x_i \sigma^x_j \rangle(t)$ converges in the long-time limit for any finite lattice. Only in the
thermodynamic limit of an infinite number of lattice sites do we have a chance of observing convergence
towards an equilibrium value. To derive such a result, we consider regular $d$-dimensional lattices. Without
loss of generality we consider the lattice constants normalised such that there is on average one lattice site
per unit (hyper)volume in the limit of large lattice size. We choose coupling constants decaying like a
power-law with the Euclidean distance $|i-j|$ between lattice sites $i$ and $j$,
\begin{equation}
J_{i,j}=\frac{J}{|i-j|^\alpha}.
\end{equation}
Without loss of generality we set $J=1$. Under these conditions and in the limit of large lattice size, we obtain the bounds
\begin{widetext}
\begin{subequations}
\begin{align}
\left|P_{i,j}^-\right| \leq \mathcal{P}_{i,j}^- &:=\frac{1}{2}
\begin{cases}
\exp\left[-C_{\alpha,d}^-\left(\frac{4\alpha|i-j|t}{\pi}\right)^2 N^{1-2(\alpha+1)/d}\right] & \text{for $0\leq\alpha<d/2-1$},\\[3mm]
\exp\left[-C_{\alpha,d}^-\left(\frac{4\alpha|i-j|t}{\pi}\right)^{d/(\alpha+1)}\right] & \text{for
$\alpha>d/2-1$},
\end{cases}\label{e:finalbound-}\\
\left|P_{i,j}^+\right| \leq  \mathcal{P}_{i,j}^+ &:=\frac{1}{2}
\begin{cases}
\exp\left[-C_{\alpha,d}^+\left(\frac{8t}{\pi}\right)^2 N^{1-2\alpha/d}\right] & \text{for $0\leq\alpha<d/2$},\\[2mm]
\exp\left[-C_{\alpha,d}^+\left(\frac{8t}{\pi}\right)^{d/\alpha}\right] & \text{for $\alpha>d/2$},
\end{cases}\label{e:finalbound+}
\end{align}
\end{subequations}
\end{widetext}
with positive constants $C_{\alpha,d}^+$ and $C_{\alpha,d}^-$ as defined in
\eqref{e:C-}. The proof of the inequalities \eqref{e:finalbound-} and \eqref{e:finalbound+}
is postponed to Sec.\ \ref{s:proof}.

\begin{figure*}
  \vspace{0pt} \includegraphics[height=0.2\linewidth]{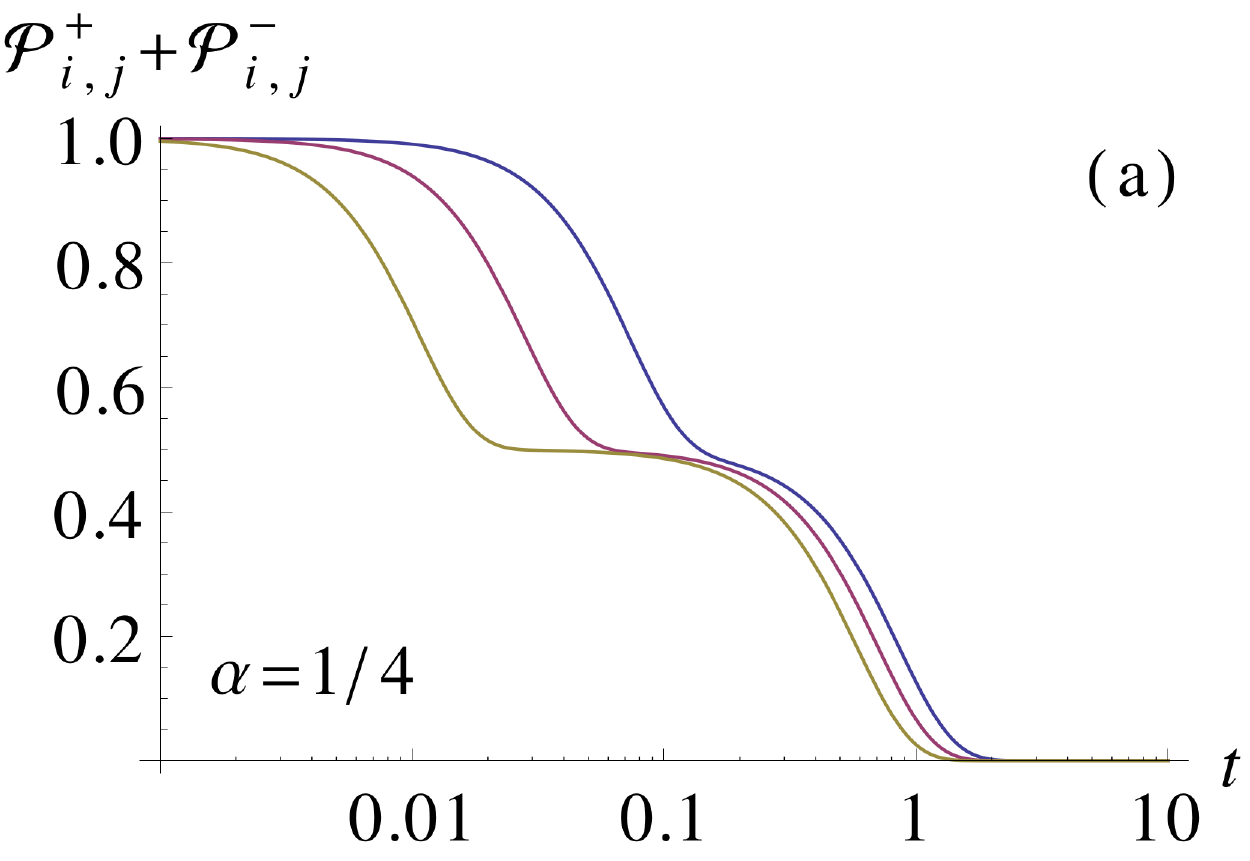}
  \hfill
  \vspace{0pt} \includegraphics[height=0.2\linewidth]{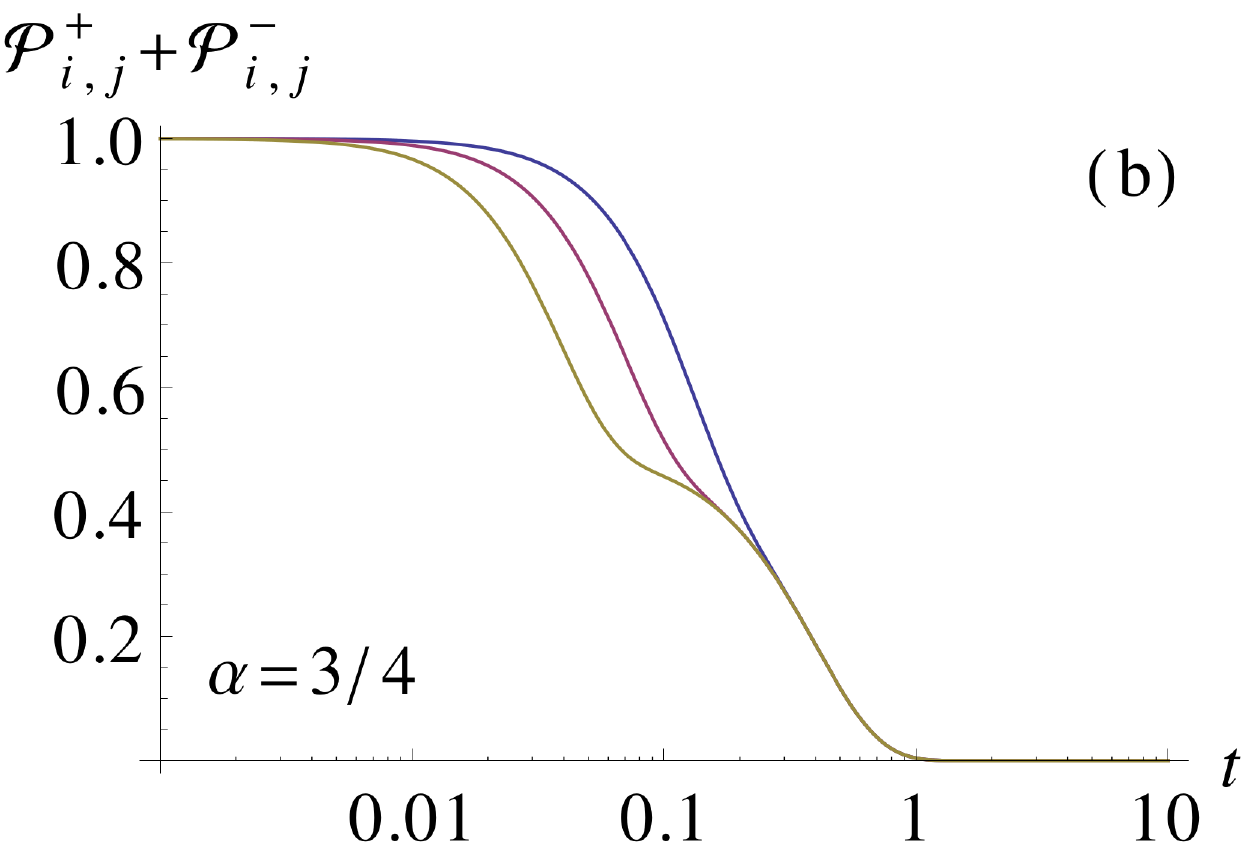}
  \hfill
  \vspace{0pt} \includegraphics[height=0.2\linewidth]{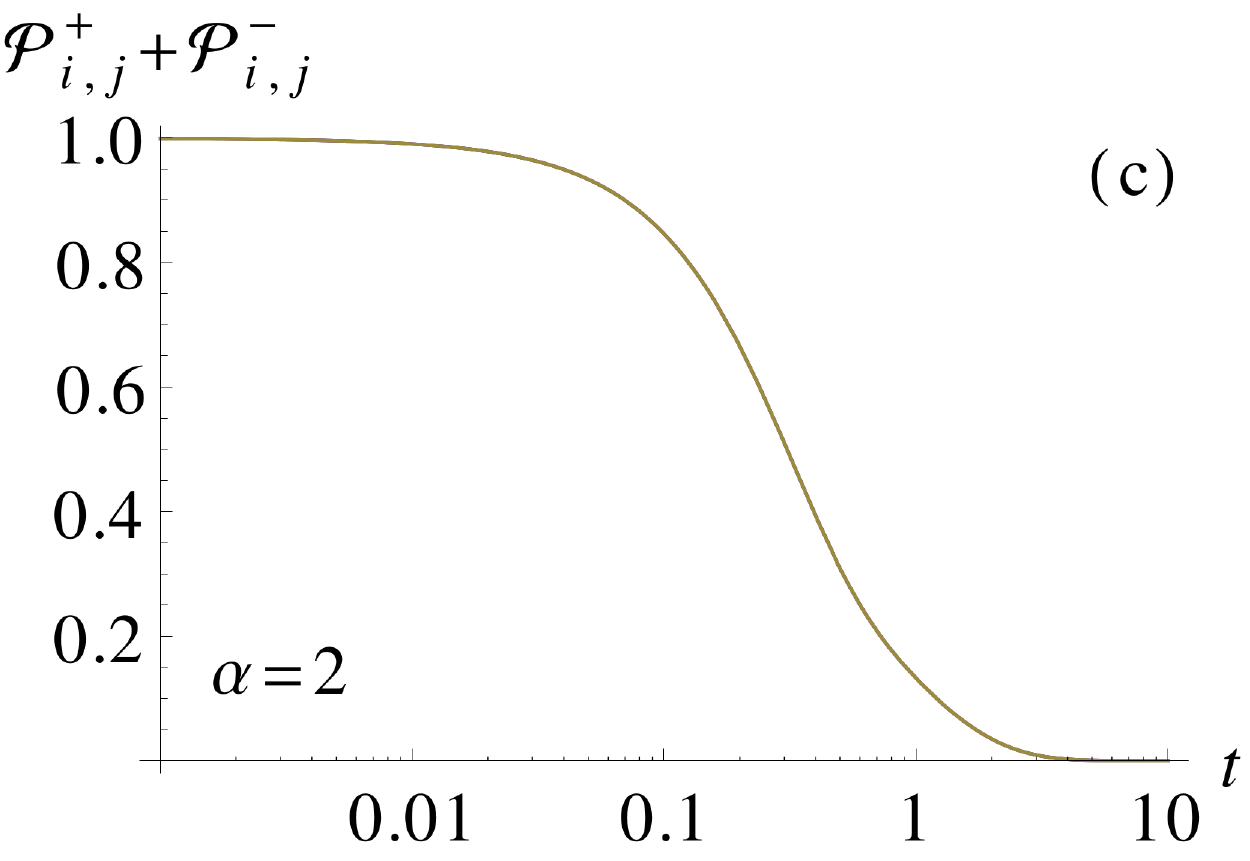}
  \hfill
  \vspace{0pt} \includegraphics[height=0.2\linewidth]{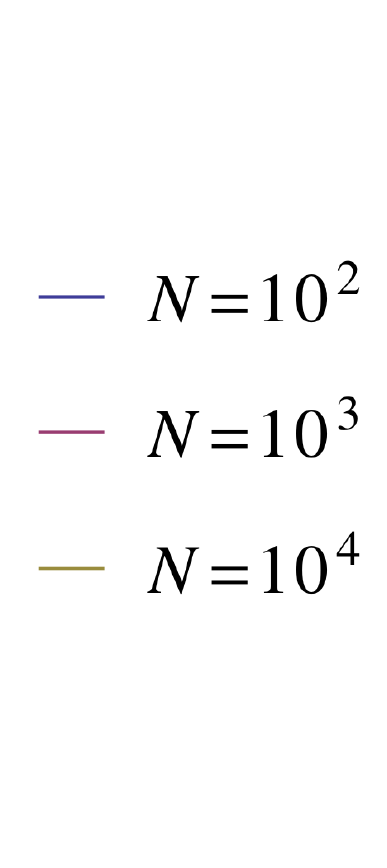}
\caption{Plots of the bound $\mathcal{P}_{i,j}^- + \mathcal{P}_{i,j}^+$ on the normalised correlator $\langle \sigma^x_i \sigma^x_j \rangle(t)/\langle \sigma^x_i \sigma^x_j \rangle(0)$ as a function of time. The examples are for lattice dimension $d=3$, and the $\alpha$-values are chosen such that they furnish examples for the three different regimes of relaxation behaviour, as discussed in the text.}
\label{fig:relaxation_dynamics}
\end{figure*}

Depending on the value of $\alpha$, the bounds \eqref{e:finalbound-} and \eqref{e:finalbound+} decay like stretched or compressed exponentials. By numerically evaluating the exact expressions \eqref{e:P}, we find that the functional form of the bound agrees well, although the numerical constants in the bound overestimate, as expected, the exact values. From Eqs.\ \eqref{e:finalbound-} and \eqref{e:finalbound+} one can read off that there are three different regimes of $\alpha$-values, each with a different relaxation or scaling behaviour:

$0\leq\alpha<d/2-1$:
$\left|P_{i,j}^-\right|$ and $\left|P_{i,j}^+\right|$ both decay like a Gaussian in time, and both do so on
time scales that are $N$-dependent. The two time scales of relaxation are widely separated, with
$\left|P_{i,j}^-\right|$ decaying much slower than $\left|P_{i,j}^+\right|$. The form of the resulting upper
bound on $\langle \sigma^x_i \sigma^x_j \rangle$ is shown in Fig.\ \ref{fig:relaxation_dynamics}a. This regime occurs for
positive $\alpha$ only in lattice dimensions $d\geq3$.

$d/2-1<\alpha<d/2$:
Again, relaxation takes place in a two-step process with widely separated time scales. The fast process
described by $\left|P_{i,j}^+\right|$ still decays like a Gaussian in time, on a time scale that is
$N$-dependent. The slow time scale corresponding to $\left|P_{i,j}^-\right|$ is independent of the system
size, with a decay in the form of a compressed exponential. The form of the resulting upper bound on $\langle
\sigma^x_i \sigma^x_j \rangle$ is shown in Fig.\ \ref{fig:relaxation_dynamics}b.

\begin{figure}[b]\centering
  \includegraphics[width=\linewidth]{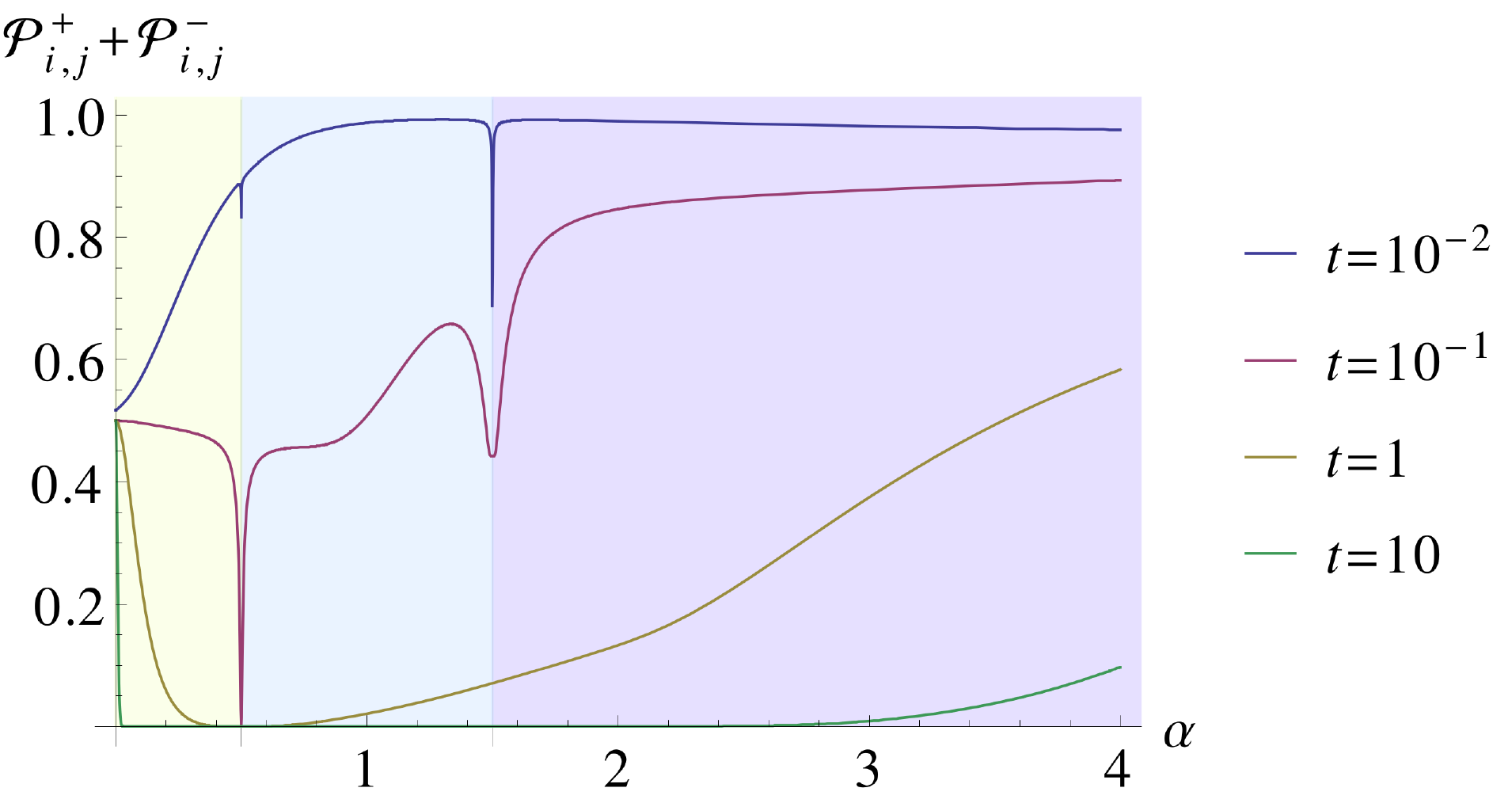}
 \caption{For fixed instances of time, the bound $\mathcal{P}_{i,j}^- + \mathcal{P}_{i,j}^+$ is shown as a
function of the exponent $\alpha$. The example is for dimension $d=3$ and a lattice of $N=10\times10\times10$
sites. The different shaded regions correspond to the three ranges of $\alpha$-values discussed in the text.}
 \label{fig:function_of_alpha}
\end{figure}

\begin{figure}[b]\centering
 \includegraphics[width=\linewidth]{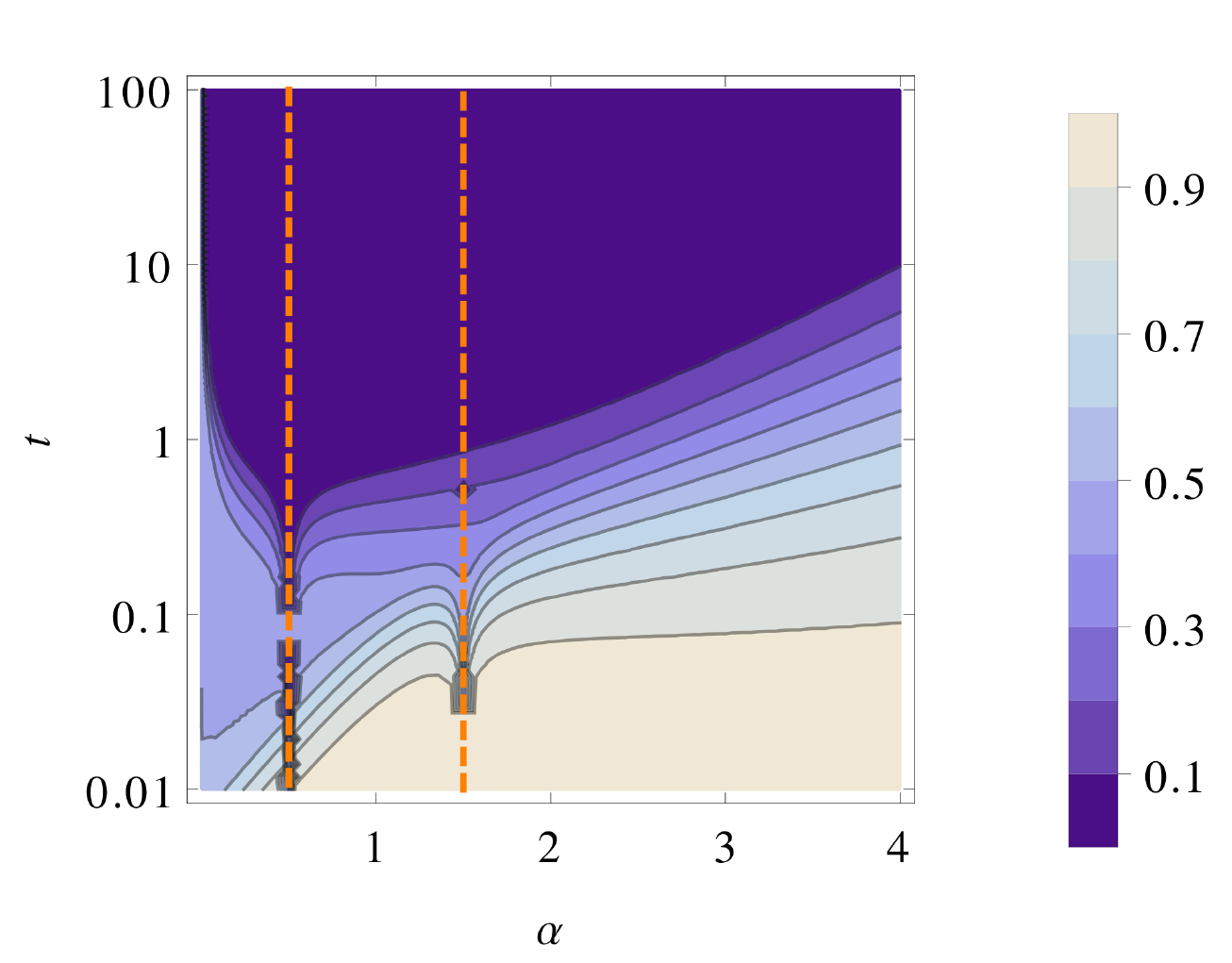} 
 \caption{Contour plot of the bound $\mathcal{P}_{i,j}^- + \mathcal{P}_{i,j}^+$ as a function of $t$ and
$\alpha$ for a three-dimensional lattice of $N=10\times10\times10$ sites. The dashed lines separate the three
regions of $\alpha$-values discussed in the text.}
 \label{fig:alpha_t_contourplot}
\end{figure}

$\alpha>d/2$:
Both terms $\left|P_{i,j}^-\right|$ and $\left|P_{i,j}^+\right|$ decay to zero like stretched or compressed
exponentials. Relaxation takes place in a single step, as both relevant time scales are very similar and
independent of $N$. The form of the resulting upper bound on $\langle \sigma^x_i \sigma^x_j \rangle$ is shown
in Fig.\ \ref{fig:relaxation_dynamics}c.

Figs.\ \ref{fig:function_of_alpha} and \ref{fig:alpha_t_contourplot} show further graphical representations of the bound, highlighting in particular the qualitative changes that occur upon variation of the exponent $\alpha$.

On the basis of these results, physical properties of the model---including dephasing dynamics, prethermalisation, and others---can be discussed similarly to the two-dimensional case reported in \cite{vdWorm_etal13}, and the reader is referred to this reference for details.

\section{Proof of Eqs.\ (6\texorpdfstring{\lowercase{a}}{a}) and (6\texorpdfstring{\lowercase{b}}{b})}
\label{s:proof}

The starting point for the derivation is the exact expression \eqref{e:P}, where $P^{\pm}_{i,j}$ is given as a
product (over all lattice sites) of cosine terms. Since $|\cos x|\leq1$, we can upper bound the absolute value
of this quantity by a product over a subset of lattice sites,
 \begin{equation}\label{e:P2}
 P^{\pm}_{i,j}\leq\frac{1}{2}\prod_{k\in\Lambda\setminus g^{\pm}_{i,j}(t)} \cos\left[2\left(J_{k,i}\pm J_{k,j}\right)t\right],
\end{equation}
where we have defined
\begin{equation}\label{constraint}
g^{\pm}_{i,j}(t):=\left\{k\in\Lambda: \left| 2\left( J_{k,i}\pm J_{k,j}\right)t \right| \geq \frac{\pi}{2} \right\}.
\end{equation}
This subset is chosen such that, for all $k\in\Lambda\setminus g^{\pm}_{i,j}(t)$, we can make use of the inequality
\begin{align}
 |\cos(\pi x)|\leq 1-4x^2\leq
\exp\left(-4x^2\right), \label{eq:inequality}
\end{align}
valid for all $|\pi x|<2$, to write
\begin{equation}\label{eq:bound_02}
 P^{\pm}_{i,j} \leq \frac{1}{2} \exp\Biggl[ -\left(\frac{2t}{\pi}\right)^2 \sum_{k\in\Lambda\setminus g^{\pm}_{i,j}(t)} \left( J_{k,i}\pm J_{k,j} \right)^2 \Biggr].
\end{equation}

\begin{figure}
\includegraphics[height=0.7\linewidth]{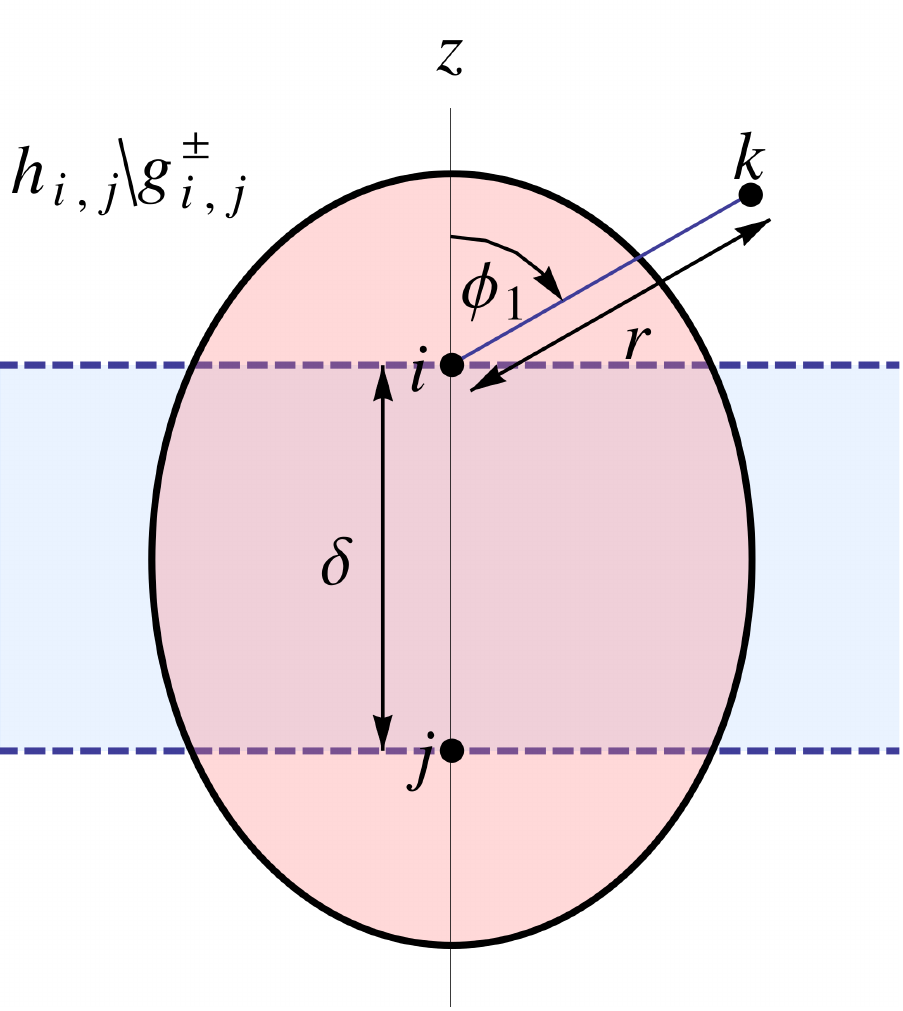}
\caption{Sketch of the regions $g_{i,j}^\pm(t)$ and $h_{i,j}$ and of the chosen coordinate system as used for
the proof in Sec.\ \ref{s:proof}.}
\label{fig:partial_ball}
\end{figure}

We restrict the $k$-summation even further by excluding the hyperslab $h_{i,j}$ sketched in Fig.\ \ref{fig:partial_ball} \footnote{This exclusion is not necessary, but it simplifies the calculation.}. The occurrence of Euclidean distances in the couplings $J_{i,k}$ and $J_{j,k}$ then suggests to parametrise the lattice sites $k\in\Lambda$ by hyperspherical coordinates,
\begin{equation}
k(r,\phi_1,\dotsc,\phi_{d-1})=
\begin{pmatrix}
 r\cos\phi_1 \\
 r\sin\phi_1\cos\phi_2\\
 \vdots\\
 r \sin\phi_1 \cdots \sin\phi_{d-2} \cos\phi_{d-1}\\
 r \sin\phi_1 \cdots \sin\phi_{d-2} \sin\phi_{d-1}
\end{pmatrix},
\end{equation}
with the origin of the coordinate system placed at lattice site $i$ and the $z$-axis chosen along the line connecting $i$ and $j$. The couplings can then be written as
\begin{equation}
J_{i,k}=r^{-\alpha},\quad J_{j,k}=\left(r^2+2r\delta\cos\phi_1+\delta^2\right)^{-\alpha/2},
\end{equation}
where $\delta=|i-j|$ denotes the distance between sites $i$ and $j$. It is then convenient to further restrict the $k$-summation in \eqref{eq:bound_02} by excluding the hyperslab sketched in Fig.\ \ref{fig:partial_ball}. Exploiting also the reflection symmetry of the problem, we arrive at the bound
\begin{equation}\label{exp}
 P^{\pm}_{i,j} \leq \frac{1}{2}\exp\Biggl( -\frac{8t^2}{\pi^2} \sum_{k\in h_{i,j}\setminus g^{\pm}_{i,j}(t)} \left( J_{k,i}\pm J_{k,j} \right)^2 \Biggr)
\end{equation}
with the $k$-summation restricted to the lattice sites in the half plane
\begin{equation}\label{eq:the_cone}
h_{i,j}=\left\{ k(r,\phi_1,\dotsc,\phi_{d-1})\in \Lambda : 0\leq\phi_1\leq\frac{\pi}{2}\right\}.
\end{equation}

For large lattices, we can bound the sum in \eqref{exp} by an integral,
\begin{multline}\label{eq:bound_03}
\sum_{k\in h_{i,j}\setminus g^{\pm}_{i,j}(t)} \left( J_{k,i}\pm J_{k,j} \right)^2
\geq 2\pi\mathcal{K}(d)\int_{R^{\pm}(t)}^{N^{1/d}}\rmd r\,r^{d-1}\\
\times \int_0^{\pi/2}\rmd\phi_1\,\cos\phi_1\sin^{d-2}\phi_1
  \left(J_{k,i}\pm J_{k,j}\right)^2,
\end{multline}
where the prefactor
\begin{equation}
 \mathcal{K}(d)=\begin{cases}
\frac{1}{\pi} & \text{for $d=2$},\\
\prod^{d-2}_{m=2}\frac{\sqrt{\pi }\Gamma\left(\frac{1}{2} (-m+d )\right)}{\Gamma\left(\frac{1}{2}(1-m+d )\right)} &\text{for $d\geq3$},
\end{cases}
\end{equation}
originates from the integrations over $\phi_2,\dots,\phi_{d-1}$.

The lower limit $R^\pm$ of the $r$-integration still needs to be determined such that the region $g^{\pm}_{i,j}$ is excluded. I.e., we need to determine $R^\pm$ such that
\begin{equation}\label{eq:Rpm_inequality}
\left|2t\left(J_{k,i}\pm J_{k,j}  \right)\right|< \frac{\pi}{2}
\end{equation}
for all $r\geq R^\pm(t)$. We are interested in the long-time asymptotic behaviour, and in this limit large values of $r$ are required to satisfy the above inequality. Hence we can assume that $r$ is much larger than $\delta$ and expand
\begin{align}
J_{k,i}\pm J_{k,j} &=\frac{1}{r^{\alpha}}\pm \frac{1}{\sqrt{\delta^2+2r\delta\cos\phi_1 +r^2}^{\,\alpha/2}}\nonumber\\
&\sim r^{-\alpha}\pm r^{-\alpha}\left(1+\frac{\alpha \delta \cos\phi_1}{r}\right)
\label{final_step}
\end{align}
to leading order in the small parameter $\delta/r$, yielding
\begin{subequations}
\begin{align}
J_{k,i}+J_{k,j} &\sim 2r^{-\alpha},\label{e:R+}\\
J_{k,i}-J_{k,j} &\sim -\frac{\alpha \delta \cos\phi_1}{r^{\alpha+1}}.\label{e:R-}
\end{align}
\end{subequations}
Inserting these asymptotic expressions into \eqref{final_step}, we obtain
\begin{equation}
R^+(t)\sim\left(\frac{8t}{\pi}\right)^{1/\alpha},\qquad R^-(t)\sim\left(\frac{4\alpha\delta t}{\pi}\right)^{1/(1+\alpha)},
\end{equation}
valid for sufficiently large $t$.

For similar reasons we can insert the expansions \eqref{e:R+} and \eqref{e:R-} into the integrand of \eqref{eq:bound_03}. The integrations become elementary in this case, yielding
\begin{widetext}
\begin{subequations}
\begin{align}
\sum_{k\in h_{i,j}\setminus g^{\pm}_{i,j}(t)} \left(J_{i,k} + J_{k,j} \right)^2
 &\geq \frac{8\pi \mathcal{K}(d) }{(d-1)(d-2\alpha)}\left[
N^{1-2\alpha/d}-\left(\frac{8t}{\pi} \right)^{d/\alpha-2}\right],\label{e:pluscase}\\
 \sum_{k\in h_{i,j}\setminus g^{\pm}_{i,j}(t)} \left(J_{i,k}- J_{k,j} \right)^2
 &\geq \frac{4\pi \mathcal{K}(d) \alpha^2 \delta^2}{(d^2-1)(d-2\alpha-2)}
\left[N^{1-2(\alpha+1)/d}-\left(\frac{4\alpha\delta t}{\pi}\right)^{d/(\alpha+1)-2} \right],\label{e:minuscase}
\end{align}
\end{subequations}
in the limit of large $N$ and $t$. Interestingly, the bound in \eqref{e:pluscase} is independent of the distance $\delta$ between the lattice sites.

Depending on the sign of the exponents $1-2\alpha/d$ and $1-2(\alpha+1)/d$ in the $N$-terms, either the first or the second term in the square brackets of \eqref{e:pluscase} and \eqref{e:minuscase} will give the dominant contribution in the limit of large system size $N$. As a result, the asymptotic behaviour of the bounds is different for different ranges of $\alpha$,
\begin{subequations}
\begin{align}
\sum_{k\in h_{i,j}\setminus g^{\pm}_{i,j}(t)} \left(J_{i,k} + J_{k,j} \right)^2
 &\geq \frac{8\pi \mathcal{K}(d) }{(d-1)(d-2\alpha)}
 \begin{cases}
 N^{1-2\alpha/d} & \text{for
$0\leq\alpha<d/2$},\\[2mm]
-\left(\frac{8t}{\pi}\right)^{d/\alpha-2} & \text{for $\alpha>d/2$},
\end{cases}\\
\sum_{k\in h_{i,j}\setminus g^{\pm}_{i,j}(t)} \left(J_{i,k}- J_{k,j} \right)^2
 &\geq \frac{4\pi \mathcal{K}(d) \alpha^2 \delta^2}{(d^2-1)(d-2\alpha-2)} 
 \begin{cases}
 N^{1-2(\alpha+1)/d} & \text{for
$0\leq\alpha<d/2-1$},\\[3mm]
-\left(\frac{4\alpha\delta t}{\pi}\right)^{d/(\alpha+1)-2} & \text{for $\alpha>d/2-1$}.
\end{cases}
\end{align}
\end{subequations}
Inserting these expressions into the inequality \eqref{eq:bound_02} and defining the positive constants
\begin{equation}
C_{\alpha,d}^+ = \frac{\pi\mathcal{K}(d)}{(d-1)|d-2\alpha|},\qquad
C_{\alpha,d}^- = \frac{2\pi\mathcal{K}(d)}{(d^2-1)|d-2\alpha-2|},\label{e:C-}
\end{equation}
the main results \eqref{e:finalbound-} and \eqref{e:finalbound+} of the paper follow.
\end{widetext}

\bibliography{LIUP}

\end{document}